\documentclass[aps,jcp,amsmath,amssymb,showpacs,letterpaper,twocolumn]{revtex4-1}%
\usepackage{graphicx}
\usepackage{amsmath}
\usepackage{graphicx}%
\usepackage{amsfonts}%
\usepackage{amssymb}
\usepackage{color}
\usepackage{setspace}
\usepackage{textcomp}
\usepackage{multirow}
\usepackage{hyperref}
\usepackage{booktabs}
\begin{document}
	
	\title{Interplay of ion availability and mobility in the loss of cation selectivity for CaCl\textsubscript{2} in negatively charged nanopores: molecular dynamics using scaled-charge models.} 
	\author{Salman Shabbir}
	\affiliation{Center for Natural Sciences, University of Pannonia, P.O. Box 158, H-8201 Veszpr\'em, Hungary}
	\affiliation{Department of Engineering, Reykjavik University, Menntavegur 1, 102 Reykjavík, Iceland}

	\author{Dezs\H{o} Boda}
	\affiliation{Center for Natural Sciences, University of Pannonia, P.O. Box 158, H-8201 Veszpr\'em, Hungary}
	
	\author{Zolt\'an Hat\'o}\email[Author for correspondence:]{hato.zoltan@mk.uni-pannon.hu}
	\affiliation{Center for Natural Sciences, University of Pannonia, P.O. Box 158, H-8201 Veszpr\'em, Hungary}
	
	\date{\today}
	
	% Keywords: activity coefficient, solvation, Grand Canonical Monte Carlo, Born energy
	
	\begin{abstract}
		Ion transport through charged nanopores is commonly interpreted in terms of the electrical double layer structure, leading to the expectation of cation-selective conduction in negatively charged pores. 
		This picture can break down for multivalent electrolytes, where strong ion-surface correlations modify transport behavior. 
		Here, we study NaCl and CaCl$_2$ conduction through negatively charged silica nanopores using atomistic molecular dynamics simulations with scaled-charge (and also full-charge) ion models. By separating concentration, $c_i(r)$, and velocity, $v_i(r)$, contributions to the radial particle current density, $j_i(r)=c_i(r)v_i(r)$, we connect static adsorption to dynamic perm-selectivity.  
		We show that strongly adsorbed, but immobilized Ca$^{2+}$ ions and low availability of Cl$^-$ ions in the surface layer near the charged wall make the contribution of this layer to the total conduction (surface conduction) small.
		It is the bulk-like electrolyte in the middle of the pore whose contribution (volume conduction) dominates the selectivity behavior of the pore (bulk-like or even slightly anion selective).
		Although this qualitative mechanism is robust, its detailed manifestation depends sensitively on the balance of ion-surface and ion-water interactions encoded in the force field.
		% \begin{center}
			% \includegraphics*[width=0.4\textwidth]{figs_scaling/toc}\\
			% TOC figure
			% \end{center}
	\end{abstract}
	
	% \pacs{02.70.-c, 02.70.Uu, 05.10.Ln}
	
	\maketitle
	
	%#######################################################################
	%#######################################################################
	
	\clearpage
	
	\section{Introduction}
	
	Ion transport through charged nanopores is governed by a subtle interplay between surface chemistry, electrostatic correlations, and hydration effects.
	While pores bearing negative surface charge are typically expected to exhibit cation selectivity, recent experiments~\cite{zheng_ep_2003,van_der_Heyden_prl_2006,li_aca_2019,li_nl_2015,lin_jacs_2020,siwy_prl_2002,siwy_cej_2006,he_jacs_2009,gillespie_bj_2008_nanopore} and simulations~\cite{lorenz_pre_2007,Bourg_2011,hartkamp_pccp_2015,dopke_jpcc_2019,wang_jpcc_2021,fabian_jml_2022,rojano_pf_2024} have shown that this picture can break down in the presence of multivalent ions.
	
	These phenomena can be analyzed from both static and dynamic perspectives.
	From the static viewpoint, cations adsorb at deprotonated surface sites, forming contact and solvent-separated ion pairs whose structural motifs may include water molecules and atoms of the pore material.~\cite{Hocine_2016,dopke_jpcc_2019,Malloggi_2019,wang_jpcc_2021,Selmani_2023}
	The resulting charge distributions and electrostatic potential profiles define the structure of the electrical double layer (EDL).
	Although the classical Gouy–Chapman model provides a useful macroscopic interpretation~\cite{hartkamp_pccp_2015,hartkamp_cocim_2018,Malloggi_2019,siboulet_jpcc_2017,wang_jpcc_2021}, molecular-level insight is most effectively obtained from statistical-mechanical approaches (particularly, molecular simulations) based on models of increasing complexity.
	
	From the dynamic viewpoint, one considers the transport of electrolyte species through the pore under an external electric field, concentration gradient, or pressure difference.
	Transport properties are accessible in experiments, whereas theoretical descriptions span continuum transport equations~\cite{cervera_epl_2005,matejczyk_jcp_2017}, molecular simulations, and hybrid approaches coupling the two~\cite{boda_jctc_2012}.
	Ionic currents depend not only on structural features, but also on the mobilities of ions and how structural features influence mobilities.
	
	Multivalent electrolytes provide an especially revealing case. 
	In CaCl$_2$ solutions, strong specific adsorption of Ca$^{2+}$ at deprotonated sites may overcompensate the surface charge, leading to charge inversion.~\cite{hartkamp_pccp_2015,hartkamp_cocim_2018,Malloggi_2019,siboulet_jpcc_2017,wang_jpcc_2021}
	Classical EDL descriptions remain valid only if specifically adsorbed ions and their associated pairs are treated explicitly as a microscopic Stern layer rather than as part of a continuous diffuse region.~\cite{Lorenz_2008,Hocine_2016,Malloggi_2019,Ma_2021}
	These adsorbed and surface-correlated ions behave effectively as immobilized charge, modifying local mobility and obscuring the distinction between bound and conducting species.
	A direct dynamical consequence is that, once charge inversion occurs, anions can be stabilized near the interface and “leak’’ through the pore along the inverted potential landscape, eroding cation perm-selectivity for 2:1 electrolytes and even reversing it for 3:1 systems.~\cite{he_jacs_2009}
	
	In this work, we aim to establish an explicit connection between static interfacial structure and dynamic transport, focusing on how pore charge and confinement influence ionic conduction (cation versus anion selectivity, in particular) at silica-electrolyte interfaces.
	While static interfacial properties have been extensively characterized~\cite{hartkamp_pccp_2015,siboulet_jpcc_2017,dopke_jpcc_2019,wang_jpcc_2021,wang_jpcc_2023}, the manner in which adsorption selectivity translates into dynamic perm-selectivity is understood qualitatively, but remains to be characterized quantitatively.
	
	We connect static and dynamic properties through 
	\begin{equation}
		\mathbf{j}_{i}(\mathbf{r})=\mathbf{v}_{i}(\mathbf{r})c_{i}(\mathbf{r}),
		\label{eq:jvc1}
	\end{equation}
	where $\mathbf{j}_{i}(\mathbf{r})$ is the particle current density profile, $\mathbf{v}_{i}(\mathbf{r})$ is the local velocity profile, and $c_{i}(\mathbf{r})$ is the concentration (particle density) profile of ionic species $i$.
	In this framework, $c_{i}(\mathbf{r})$ is primarily determined by local interactions and describes the \textit{availability} of charge carriers, whereas $\mathbf{v}_{i}(\mathbf{r})$ represents their local \textit{mobility} under the effect of the applied electric field.
	
	This decomposition allows us to disentangle structural and dynamical contributions to ion conduction. 
	By partitioning the pore cross section into a surface-dominated EDL region adjacent to the wall and a more bulk-like interior region, we quantify how each domain contributes to the total ionic current and to overall perm-selectivity. 
	In particular, we analyze how these contributions differ between NaCl and CaCl$_2$ solutions, where multivalent adsorption, charge inversion, and interfacial immobilization can fundamentally alter both the spatial distribution of charge carriers and their effective mobility.
	
	Building on our earlier work~\cite{salman_jml_2025}, which established guidelines for reliable FF selection and model validation in bulk CaCl$_2$ solutions, we now extend our focus to the behavior of CaCl$_2$ in nanoscopically confined environments. 
	Silica nanopores provide a prototypical system for probing ion transport under extreme spatial restriction and surface interaction effects. 
	In such geometries, ion migration and selectivity are governed not only by bulk electrolyte properties but also by the specific interactions with the charged pore walls.
	
	Our previous study~\cite{salman_jml_2025} demonstrated that the balance between ion–ion (II) and ion–water (IW) interactions controls the transport properties of aqueous CaCl$_2$ in bulk, and that careful tuning of FF parameters (particularly, parallel ionic charge scaling and diameter reduction) can mitigate unphysical slow dynamics while improving agreement with experimental diffusion coefficients and conductivities. 
	Our systematic evaluation identified a scaled-charge ionic model (ECCR2), combined with the TIP4P/2005 water model, as providing a balanced description of transport properties in line with the findings of Martinek et al.~\cite{martinek_jcp_2018} for structural (neutron scattering), and viscosity properties.
	
	Note that the balance of interactions between species carrying charges (ions, water, surface groups) is also present in the nanopore system, but the competition for Ca$^{2+}$ is now involves the surface groups as a new competitor.
	While in the bulk the competition of Cl$^{-}$ ions and water molecules at the Ca$^{2+}$ ions was the main balancing process, in the pore it is rather the competition of Ca$^{2+}$ and water at the charged surface groups.
	
	In the present work, we apply the ECCR2+TIP4P/2005 combination, alongside representative full-charge and other scaled-charge models, to investigate the conduction of NaCl and CaCl$_2$ solutions in silica nanopores using atomistic molecular dynamics (MD) simulations. 
	Our objective is to elucidate how confinement and interfacial chemistry reshape ion distributions, mobilities, and ultimately perm-selectivity, and to assess to what extent FF choices that perform well in bulk remain transferable under strong confinement.
	
	\section{Models and method}
	
	\subsection{Model of the electrolyte}
	
	In our previous paper\cite{salman_jml_2025}, we tested various force fields (FF) and concluded that models that scale the charges of the ions better describe the dynamics of ions in the electrolyte. 
	In these models the charges of the ions (and charged particles) are divided by the high-frequency dielectric constant as $ q_{\text{scaled}} = q_{\text{original}}/\sqrt{\epsilon_{\infty}} $ to take into account electronic polarization and charge transfer.~\cite{leontyev_jcp_2009,leontyev_pccp_2011}
	
	This electronic continuum correction (ECC) approach was used in several works that focused on the dynamics of the system~\cite{kann_jcp_2014,biriukov_pccp_2018,yue_mp_2019,kohagen_jpcb_2014,martinek_jcp_2018,zeron_jcp_2019,duboue_jcp_2020,predota_jml_2020} or on structural features based on comparison with neutron scattering data.~\cite{kohagen_jpcb_2014,martinek_jcp_2018,megyes_jml_2006}
	If thermodynamic properties are targeted, compensation is needed for the full charges~\cite{biriukov_jcp_2022,young_jced_2019}.
	
	\begin{table*}[t!]
		\begin{center}
			\caption{Distance and charge parameters of the ionic FF parameters~\cite{kohagen_jpcb_2014,martinek_jcp_2018,oostenbrink_jcc_2004,dang_jcp_1995,kohagen_jpcb_2015}.}
			\label{tab1}
			\begin{tabular}{l|l|l|l|l}
				\hline
				\hline
				& $d_+$ / nm  & $q_+ /e$ & $d_-$ / nm & $q_- / e $  \\
				\hline
				\hline
				& \multicolumn{2}{c|}{Na$^{+}$} & \multicolumn{2}{c}{Cl$^{-}$} \\
				\hline
				ECCR-like & 0.2115 & 0.75 & 0.41 & -0.75 \\
				\hline
				& \multicolumn{2}{c|}{Ca$^{2+}$} & \multicolumn{2}{c}{Cl$^{-}$} \\
				\hline
				FULL & 0.28196  & 2 &   0.44499 &  -1 \\
				ECC & 0.28196  & 1.5 &  0.44499 &  -0.75 \\
				ECCR2 & 0.26656  & 1.5 &  0.41 &  -0.75 \\
				ECCR & 0.25376  & 1.5 &  0.37824 &  -0.75 \\
				\hline
				& \multicolumn{2}{c|}{} & \multicolumn{2}{c}{O$ _{\mathrm{S}} $} \\
				\hline
				Full O$ _{\mathrm{S}} $ &  &  & 0.307 & -0.74 \\
				Scaled O$ _{\mathrm{S}} $ &  &  & 0.307 & -0.555 \\
				\hline
				\hline
			\end{tabular}
		\end{center}
	\end{table*}
	
	While there are plenty of models in the literature for CaCl$ _{2} $,\cite{mackerell_jpcb_1998,oostenbrink_jcc_2004,dang_jcp_1995,deublein_jpcb_2012,mamatkulov_jcp_2013,young_jced_2019,zeron_jcp_2019} we focused on those developed by the group of Pavel Jungwirth.~\cite{kohagen_jpcb_2014,martinek_jcp_2018}
	The starting full-charge model (FULL) from which the scaled-charge models were developed uses the GROMOS 53a6 FF parameters from Ref.~\onlinecite{oostenbrink_jcc_2004} for Ca$^{2+}$ and from Ref.~\onlinecite{dang_jcp_1995} for Cl$^{-}$.
	
	Dividing by $\sqrt{\epsilon_{\infty}}$ is equivalent to multiplying by $0.75$ in an aqueous electrolyte; this leads to the ECC model. 
	In the ECC model, the diameters of the ions were kept at the values of the FULL model: $d_{2+}=0.282$ nm for Ca$^{2+}$ and $d_{-}=0.445$ nm for Cl$^{-}$ (see Table \ref*{tab1}).
	Kohagen et al.\cite{kohagen_jpcb_2014}, however, realized that good agreement with experiments (neutron scattering, diffusion constant, viscosity) can be achieved only if the ionic diameters are reduced. 
	This realization has led to the ECCR2 ($d_{2+}=0.267$ nm and $d_{-}=0.41$ nm~\cite{martinek_jcp_2018}) and ECCR ($d_{2+}=0.254$ nm and $d_{-}=0.378$ nm~\cite{kohagen_jpcb_2014}) models.
	
	In our previous work~\cite{salman_jml_2025}, we found the best agreement with experimental diffusion constant and specific conductivity data for the ECCR2 model (used together with the TIP4P/2005 water model), because the II and IW interactions seem to be balanced in a way that is the closest to reality.  
	Scaled-charge FFs have also been used in confined geometries to simulate ionic transport, and to study the binding of ions to charged structural groups on the surface.~\cite{vazdar_jpcb_2013,kohagen_jpcl_2014,melcrova_sr_2016,magarkar_jpcl_2017,biriukov_pccp_2018,biriukov_jpcc_2020,lebreton_jcp_2020}
	
	\subsection{Pore model}
	
	In this work, we place the electrolyte in a narrow negatively charged silica nanopore and study how ionic transport (especially, cation vs.\ anion selectivity) depends on the electrolyte and pore models. 
	We built this pore using a special tool called PoreMS. \cite{PoreMS_Kraus_2021}
	This Python package helps generate pore structures and models of such channels for use in molecular simulations.
	
	In the first step, using PoreMS, we created a silica block with dimensions of $7 \times 7 \times 7$ nm. %#TODO 
	Then, the pore was carved out and the protonation states of the dangling Si--O groups were determined. 
	By explicitly setting the fraction of protonated versus deprotonated silanol groups on the surface in the model, the surface charge can be tuned to mimic experimental values, capturing the essential chemistry of silica nanopores in aqueous environments. 
	
	In our model, we placed $104$ negatively charged silanol groups on the surface of the pore ($97$ isolated and $7$ geminal oxygens).
	The average distance of the oxygen atoms of the groups (O$_{\mathrm{S}}$) from the pore axis is $\approx 1.3$ nm over a pore length of $\approx 7$ nm.
	We define the approximate pore radius with this value.
	This arrangement corresponds to $\approx 1.54$/nm$^2$ surface group density, which corresponds to $\sigma \approx -1.54$ $e$/nm$^2$ surface charge density if we attribute $-e$ charge to each O$_{\mathrm{S}}$ atom.  
	For pores with different radii, other values for the number of silanol groups are used, but their surface density is retained.
	
	PoreMS, however, assigns charges $-0.74e$ to the O$_{\mathrm{S}}$ atoms.
	That value implies a surface charge density $\approx -1.14$ $e$/nm$^{2}$.
	Furthermore, in this study, we scale the charges of the O$_{\mathrm{S}}$ atoms (resulting in $-0.555e$ charges) to treat them consistently with the ECC models of ions, which yields a surface charge density $\approx -0.85$ $e$/nm$^{2}$ (the number of surface groups is unchanged for a given pore radius).
	
	For comparison, Gulmen and Thompson~\cite{gulmen_l_2006} also used the value $-0.74e$ for the O$_{\mathrm{S}}$ charge.
	Their FF is based on that of Br\'{o}dka and Zerda~\cite{brodka_jcp_1996}, whose value is $-0.533e$.
	These values are in accordance with those produced by PoreMS and scaled by 25\%.
	While these values are scaled, in the series of papers by Hartkamp, Siboulet,  Dufr\^eche and coworkers~\cite{hartkamp_pccp_2015,siboulet_jpcc_2017,dopke_jpcc_2019,wang_jpcc_2021,wang_jpcc_2023}, the full $-e$ charge was used for the silanol oxygens. 
	
	All these surface charge densities are realistic at large pH values.
	In their MD simulations to study ion adsorption on silica surfaces, for example, Wang et al.~\cite{wang_jpcc_2021} used silanol group density $2.95$/nm$^{2}$, a value typical for amorphous silica surfaces~\cite{zhuravlev_l_1987,luo_jrs_2001}. 
	PET nanopores in the experiments of Siwy et al.~\cite{he_jacs_2009,gillespie_bj_2008_nanopore} typically carry $\approx -1$ $e$/nm$^{2}$ surface charge.
	P{\v{r}}edota et al.~\cite{predota_l_2016} used values up to $-2.5$ $e$/nm$^{2}$ for the rutile surface charge in their MD study for the origin of the zeta potential.
	In general, surface charge densities depend on pH and the degree of the overlap of EDLs~\cite{yang_pccp_2020}.
	
	As most of the simulation works in literature, we use a fixed number of charges on the pore wall and neglect fluctuations due to surface protonation equilibria.
	After building the silica pore, we attached two boxes of aqueous electrolyte to both ends of the silica block. 
	We apply periodic boundary conditions resulting in periodic images of the system including the membrane and the nanopore in every spatial direction.
	Since the pore had net negative charge, we added extra cations to make the whole simulation cell charge neutral.
	Once built with PoreMS, the atoms of the silica block were fixed but allowed to vibrate, constrained by a harmonic potential.
	We use this simple approach to take surface flexibility into account partially.
	
	\subsection{Molecular dynamics}
	
	MD simulations have been performed using the GROMACS molecular simulation software suite v.2023.2.\cite{hess_jctc_2008}. 
	The simulation cell was a rectangle with a length of $\approx 18$ nm and contained a $1$ M CaCl$_{2}$ solution. 
	Depending on the FF, this means 373--381 $\mathrm{Ca}^{2+}$ and 695--730 $\mathrm{Cl}^{-}$ ions with 18500--18900 water molecules (for $R^{\mathrm{P}}\approx 1.3$ nm).
	The LINCS algorithm~\cite{hess_jcc_1997} was employed to keep the water molecules rigid by correcting the positions of oxygen and hydrogen atoms after a time step in which they are allowed to change.
	
	Short-range interactions, such as electrostatic (Coulomb) and van der Waals forces (12-6 Lennard-Jones), were adjusted using a cutoff of $1.2$ nm using the Verlet cutoff approach. 
	Long-range electrostatic interactions were computed using the Particle Mesh Ewald method~\cite{essmann_jcp_1995}.
	The simulation cell is sufficiently large so that the interaction between periodic images becomes weak and the resulting artifacts are generally small.
	The temperature was maintained at $298.15$ K with a $v$-rescale (Berendsen-type) thermostat with a coupling constant of $0.5$ ps.\cite{bussi_jcp_2007} 
	The Parrinello-Rahman barostat~\cite{parrinello_jap_1981} maintained a pressure of $1$ bar by adjusting the system's volume to ensure stability (coupling constant $0.5$ ps). 
	
	First, we performed an energy minimization step, getting rid of any awkward or too-close contacts between atoms that might cause problems. 
	After that, we ran an equilibration simulation under $NpT$ conditions for about $10$ nanoseconds (no external electric field applied here). 
	This step was crucial for pushing water and ions into the channel, making sure it was completely filled and there were no empty spaces left. 
	The production runs were performed in the presence of an electric field in the $NVT$ ensemble.
	The direction of the electric field was along the $z$-axis (the rotational axis of the pore), while its magnitude was $E=0.06606$ V/nm.
	By running simulations for varying electric field strengths, we show in the \hyperref[si]{Supplementary Material} (\hyperref[si]{\hyperref[si]{SM}}) that this value falls into the linear response regime.
	We employed the leap-frog integrator with a time step of $1$ fs.
	Data for positions and velocities were collected every $5$ ps for post-process analyses during a $100$ ns long production run.
	Time dependence of profiles and integrated current values are shown in the \hyperref[si]{SM} to demonstrate that the simulations provide converged results during this time.
	
	\subsection{Simulated quantities}
	
	Since the electric field has only a $z$ component ($E_{z}=E$), the relevant component of the particle number current density, $\mathbf{j}_{i}(\mathbf{r})$, and the velocity, $\mathbf{v}_{i}(\mathbf{r})$, is the $z$ component, $j_{i,z}(\mathbf{r})$ and $v_{i,z}(\mathbf{r})$ for ionic species $i$, that we will denote with $j_{i}(\mathbf{r})$ and $v_{i}(\mathbf{r})$ for concise notation.
	
	Although the pore surface is rough, and the charged groups on the wall are localized and distributed only approximately uniformly, we average over the azimuthal angle, $\phi$.
	Furthermore, we focus on the behavior of the profiles inside the pore, where we assume that the main determinant of the inhomogeneity is the surface charge on the pore wall, which exerts its effect primarily in the radial direction.
	Consequently, we average over the axial dimension, $z$, of the pore as well.
	
	As a result, we report only the radial behavior of the profiles that are related through $j_{i}(r)=v_{i}(r)c_{i}(r)$, where $r$ is the distance from the rotational axis of the pore. 
	This equation is informative because the current density is obtained as a product of two terms of which $c_{i}$ characterizes the \textit{availability} of charge carriers, while $v_{i}$ characterizes the \textit{mobility} of charge carriers. 
	The transport properties are determined by both.
	
	To obtain the $\phi$- and $z$-averaged radial profiles, the local quantities need to be calculated first.
	Simulation data are collected for small elementary cells $\alpha$ that are curved wedge-shaped boxes defined by the outer and inner radii $r_{\mathrm{o}}^{\alpha}$ and $r_{\mathrm{i}}^{\alpha}$ (so $\Delta r^{\alpha}=r_{\mathrm{o}}^{\alpha}-r_{\mathrm{i}}^{\alpha}$), respectively, by edge length $\Delta z$ in the axial direction, and by the azimuthal angle increment  $\Delta \phi$. 
	The volume element is given as 
		\begin{equation}
			V^{\alpha}=A^{\alpha}\Delta z= \frac{1}{2} \left[ (r_{\mathrm{o}}^{\alpha})^{2}-(r_{\mathrm{i}}^{\alpha})^{2} \right] \Delta \phi \, \Delta z ,
		\end{equation}
	where $A^{\alpha}$ is the surface area perpendicular to the $z$ axis. 
	The velocity profile for the volume element $\alpha$ has been computed from
		\begin{equation}
			v_{i}^{\alpha}=\dfrac{1}{N_{i}^{\alpha}}\sum_{k=1}^{N_{i}^{\alpha}}\dfrac{\Delta z_{i,k}}{\Delta t},
			\label{eq:vialpha}
		\end{equation}
	where $N_{i}^{\alpha}\ne 0$ is the number of ionic species $i$ found in subvolume $\alpha$ during the simulation (at least, in the configurations saved by GROMACS) and $\Delta z_{i,k}$ is the displacement of an ion of species $i$ in the $z$ direction during $\Delta t$ which is the time interval for saving configurations.
	The concentration profile is computed from
		\begin{equation}
			c_{i}^{\alpha}=\dfrac{N_{i}^{\alpha}}{N_{t}V^{\alpha}},
			\label{eq:cialpha}
		\end{equation}
	where $N_{t}$ is the number of snapshots during the simulation saved in $\Delta t$ time intervals.
	If the volume element is small, Eq.~\ref{eq:jvc1} for the local quantities holds and the current density profile is obtained as the product of $v_{i}^{\alpha}$ and $c_{i}^{\alpha}$ as
		\begin{equation}
			j_{i}^{\alpha}= v_{i}^{\alpha}c_{i}^{\alpha} =\dfrac{1}{N_{t}V^{\alpha}}\sum_{k=1}^{N_{i}^{\alpha}}\dfrac{\Delta z_{i,k}}{\Delta t}.
			\label{eq:jialpha}
		\end{equation}
	The averaging process with which the radial profiles are obtained from the local profiles in Eqs.~\ref{eq:vialpha}-\ref{eq:jialpha} is described in Appendix \ref{appendix}.
	The derivation in Appendix \ref{appendix} shows that the $j_{i}(r)=c_{i}(r)v_{i}(r)$ relation holds for the averaged profiles as well.
		
	The particle current, $J_{i}$, can be computed by integrating the current density over the pore’s cross section.
	Alternatively, it can be obtained by counting ion crossings through predefined planes.
	The currents calculated from the two methods agree well.
	Charge current is defined as $I_{i}=q_{i}J_{i}$.
	
	The ratio of charge currents for cations and anions is defined as cation selectivity of the pore (or, more briefly, pore selectivity):
	\begin{equation}
		S_{+}^{\mathrm{P}}=\dfrac{I_{+}^{\mathrm{P}}}{I_{-}^{\mathrm{P}}},
	\end{equation}
	where $I_{i}^{\mathrm{P}}$ is the charge current of ion species $i$ measured in a pore simulation.
	Similarly, we define 
	\begin{equation}
		S_{+}^{\mathrm{B}}=\dfrac{I_{+}^{\mathrm{B}}}{I_{-}^{\mathrm{B}}},
	\end{equation}
	where $I_{i}^{\mathrm{B}}$ are the corresponding charge currents obtained from a bulk simulation (taken from our previous study~\cite{salman_jml_2025}). 
	
	Although it is not usual to call a bulk electrolyte selective, different diffusion constants (different interactions with water) lead to different conductivities in bulk.
	Therefore,  for short, we may call $S_{+}^{\mathrm{B}}$ ``bulk selectivity'' and in practice it means bulk-like behavior from the point of view of selectivity.
	The purpose of introducing this quantity is to normalize pore selectivity; that is, to express the ability of the pore to distinguish between Ca$^{2+}$ and Cl$^{-}$ beyond their differing mobilities in bulk. 
		This is characterized by the ratio
	\begin{equation}
		S_{+}^{*}=\dfrac{S_{+}^{\mathrm{P}}}{S_{+}^{\mathrm{B}}}.
	\end{equation}
	If $S_{+}^{*}\gg 1$, the pore is strongly cation selective, whereas if $S_{+}^{*}\sim 1$, the pore exhibits bulk-like behavior.
	
	Radial distribution functions (RDF) are defined by 
	\begin{equation}
		g_{ij}(r)=\dfrac{\rho_{ij}(r)}{\rho_{ij}^{\mathrm{cut}}} 
		,
	\end{equation}
	where $ \rho_{ij}(r)$ is the density of ions of species $j$ at distance $r$ from ions of species $i$ (or, vice versa) and $\rho_{ij}^{\mathrm{cut}}$ is the average density within an $R^{\mathrm{cut}}$ cut-off radius ($0.8$ nm, in this work).
	This normalization ensures that the same geometric error (arising from the fact that ions exist only on one side of the pore wall) is introduced in both densities. 
	As a result, the ratio remains meaningful even though the accessible volume is smaller than in the bulk.
	In practice, the RDF is computed as
	\begin{equation}
		g_{ij}(r_{k}) = \dfrac{\langle N_{ij}(r_{k}) \rangle/ V^{\mathrm{shell}}_{k}}{\langle N_{ij}^{\mathrm{cut}}\rangle/V^{\mathrm{cut}}},
	\end{equation} 
	where $\langle N_{ij}  (r_{k}) \rangle $ is the average number of ions $j$ in the $k$th shell (of width $\Delta r$) around ions $i$ (the O$_{\mathrm{S}}$ atoms, here), $V_{k}^{\mathrm{shell}}$ is the volume of the shell, $\langle N_{ij}^{\mathrm{cut}}\rangle $ is the average number of ions $j$ in the $R^{\mathrm{cut}}$ sphere around ions $i$, and $V^{\mathrm{cut}}=4\pi (R^{\mathrm{cut}})^{3}/3$ is the volume of the sphere.
	With this definition, $g_{ij}(r)$ smoothly approaches unity as $r \to R^{\mathrm{cut}}$.
		
	\section{Results and Discussion}
	
	To make full use of Eq.~\ref{eq:jvc1}, we present figures showing the radial profiles $j_{i}(r)$, $v_{i}(r)$, and $c_{i}(r)$.
	In the figures, the concentration profiles are expressed in mol/dm$^{3}$ to facilitate physical interpretation.
	In the figures we indicate whether the results are shown for the full-charged O$_{\mathrm{S}}$ model ($-0.74e$) or the scaled-charge O$_{\mathrm{S}}$ model ($-0.55e$) by ``O$_{\mathrm{S}}$ Full'' and ``O$_{\mathrm{S}}$ Scaled'', respectively.
	
	\subsection{Comparison of NaCl and CaCl\textsubscript{2}}
	
	\begin{figure*}[t!]
		\begin{center}
			\includegraphics*[width=0.9\textwidth]{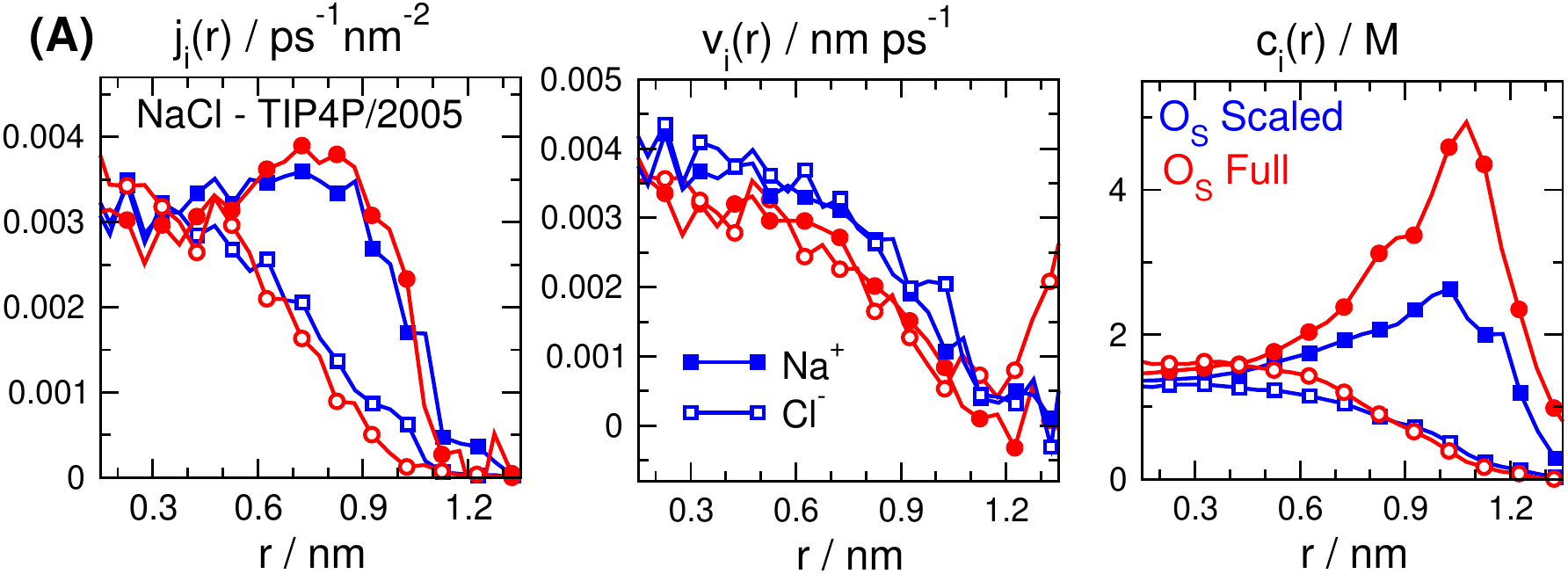}\\ \vspace{0.4cm}
			\includegraphics*[width=0.9\textwidth]{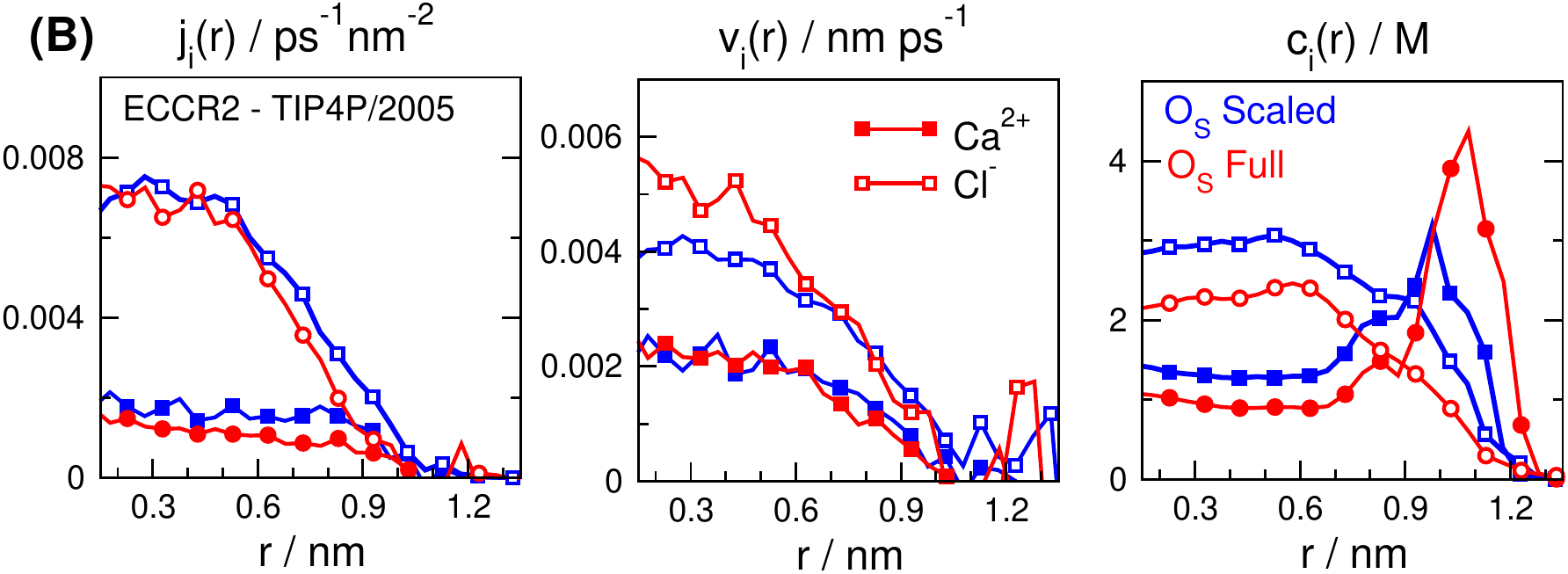}
		\end{center}
		\caption{From left to right: radial dependence of the axial ($z$) components of particle current density (in 1/ps$\;$nm$^{2}$), velocity (in nm/ps), and concentration (in mol/dm$^{3}$).
		Red and blue curves refer to full-charge and scaled-charge silanol oxygen (O$_{\mathrm{S}}$) models, respectively.
		Full and open symbols refer to cations and Cl$^{-}$ ions, respectively.
		Top row (A) refers to simulations for a scaled-charge model for NaCl with TIP4P/2005 water model \cite{kohagen_jpcb_2015}, while the bottom row (B) refers to simulations for the ECCR2 model for CaCl$_{2}$ with TIP4P/2005 water model. \cite{martinek_jcp_2018}
		}
		\label{fig1}
	\end{figure*} 
	
	The main message of this work follows from the comparison of Figs.~\ref{fig1}A and \ref{fig1}B.
	Figure \ref{fig1}A presents profiles for NaCl, while Fig.~\ref{fig1}B shows the profiles for CaCl$_{2}$.
	The difference between the two systems is striking. 
	
	For NaCl, the $c_{i}(r)$ profile exhibits a classical EDL behavior with a diffuse layer of Na$^{+}$ excess near the pore wall.
	It is usual to divide the pore interior along the radial dimension ($r$) into a surface and a volume (bulk-like) region.
	The surface region corresponds to the EDL region near the pore wall, where the negatively charged wall creates a cation selective region that is selective in both the static sense of the word (\textit{availability}: which ion is present in larger quantity in the region) and the dynamic sense of the word (perm-selectivity: which ion permeates in larger quantity through the region).
	The width of the surface region is associated (among other parameters) with the screening length of the electrolyte, which is well approximated by the Debye-length.
	The volume region is approximately charge-neutral and exhibits bulk-like behavior.
	Surface and volume conductances associated with these regions could also be defined.  
	
	Toward the wall, the Na$^{+}$ velocity decreases due to the increasing electrostatic attraction to the surface groups, while Cl$^{-}$ velocity also decreases due to the attractive interactions with the excess Na$^{+}$ ions that migrate in the opposite direction.
	The product $v_{i}(r)c_{i}(r)$ yields the $j_{i}(r)$ current density profile, whose cross section integral is the current. 
	The left hand panel of Fig.~\ref{fig1}A implies larger current for Na$^{+}$ than for Cl$^{-}$.
	The pore selectivity for NaCl is $S_{+}^{\mathrm{P}}=1.986$ for the scaled-charge O$_{\mathrm{S}}$ groups and $2.44$ for the full-charge O$_{\mathrm{S}}$ groups (numerical values with error bars are found in Table \ref{tab2}).
	The latter is larger because the total charge of the O$_{\mathrm{S}}$ groups is larger, so the Na$^{+}$ excess is larger.
	For comparison, ``bulk selectivity'' is $S_{+}^{\mathrm{B}}=0.83$ (slightly Cl$^{-}$ selective) due to the larger \textit{\textit{mobility}} (diffusion constant) of Cl$^{-}$ ions.    
	
	\begin{table*}[t!]
		\begin{center}
			\caption{Simulated cation selectivities in bulk ($S_{+}^{\mathrm{B}}$), pore ($S_{+}^{\mathrm{P}}$), and the ratio ($S_{+}^{*}$) for various ion, water, and surface oxygen models. The rows from FULL to ECC refer to CaCl$_{2}$. The pore radius is $R^{\mathrm{P}}\approx 1.3$ except for the ECCR2\,+\,TIP4P/2005\,+\,Scaled O$_{\mathrm{S}}$ case, for which pore radius dependence was simulated. The statistical uncertainties in the last two digits are shown in parentheses.}
			\label{tab2}
			\setlength{\tabcolsep}{8pt} \renewcommand{\arraystretch}{1.2}
			\begin{tabular}{l|l|l|l|l|l|l}
				\hline
				\hline
				Ion model 				& Water model 					& $S_{+}^{\mathrm{B}}$ ~\onlinecite{salman_jml_2025}			& O$_{\mathrm{S}}$ model& $R^{\mathrm{P}}$      & $S_{+}^{\mathrm{P}}$ & $S_{+}^{*}$	\\ 
				\hline \hline
				\multirow{2}{*}{NaCl} & \multirow{2}{*}{TIP4P/2005} & \multirow{2}{*}{0.830(14)} & Full & \multirow{2}{*}{1.3} & 2.44(15) & 2.94(22) \\
				& & & Scaled & & 1.986(95) & 2.39(15) \\
				\hline \hline		
				\multirow{4}{*}{\shortstack{CaCl$_2$\\FULL}} & \multirow{2}{*}{TIP4P/2005} & \multirow{2}{*}{0.645(22)} & Full & \multirow{4}{*}{1.3} & 0.246(37) & 0.381(70) \\
				& & & Scaled & & 0.613(38) & 0.950(91) \\
				\cline{2-4}\cline{6-7}
				& \multirow{2}{*}{SPC/E} & \multirow{2}{*}{0.634(19)} & Full &  & 0.755(31) & 1.191(85) \\
				& & & Scaled & & 1.039(45) & 1.64(12) \\
				\hline \hline
				\multirow{4}{*}{\shortstack{CaCl$_2$\\ECC}} & \multirow{2}{*}{TIP4P/2005} & \multirow{2}{*}{0.681(24)} & Full & \multirow{4}{*}{1.3} & 0.504(28) & 0.741(67) \\
				& & & Scaled & & 0.641(29) & 0.941(75) \\
				\cline{2-4}\cline{6-7}
				& \multirow{2}{*}{SPC/E} & \multirow{2}{*}{0.671(15)} & Full &  & 0.797(29) & 1.188(70) \\
				& & & Scaled & & 1.262(48) & 1.88(11) \\
				\hline \hline
				\multirow{7}{*}{\shortstack{CaCl$_2$\\ECCR2}} & \multirow{5}{*}{TIP4P/2005} & \multirow{5}{*}{0.713(37)} & Full & 1.3 & 0.446(27) & 0.626(71) \\
				\cline{4-7}
				& & & \multirow{4}{*}{Scaled} & 1 & 0.78(18) & 1.09(30) \\
				& & & & 1.3 & 0.658(28) & 0.923(87) \\
				& & & & 2 & 0.697(13) & 0.978(69) \\
				& & & & 3 & 0.6947(68) & 0.974(60) \\
				\cline{2-7} 
				& \multirow{2}{*}{SPC/E} & \multirow{2}{*}{0.663(17)} & Full & 1.3 & 0.857(37) & 1.293(89) \\
				& & & Scaled & 1.3 & 1.686(67) & 2.54(17) \\
				\hline \hline
				\multirow{4}{*}{\shortstack{CaCl$_2$\\ECCR}} & \multirow{2}{*}{TIP4P/2005} & \multirow{2}{*}{0.691(19)} & Full & \multirow{4}{*}{1.3} & 0.567(59) & 0.82(11) \\
				& & & Scaled & & 1.137(81) & 1.65(16) \\
				\cline{2-4}\cline{6-7}
				& \multirow{2}{*}{SPC/E} & \multirow{2}{*}{0.679(21)} & Full &  & 0.841(52) & 1.24(11) \\
				& & & Scaled & & 1.325(57) & 1.95(14) \\
				\hline \hline
			\end{tabular}
		\end{center}
	\end{table*}

	The behavior of the CaCl$_{2}$ electrolyte in the negatively charged pore is different.
	While the excess cation layer near the wall is present here as well, Ca$^{2+}$ ions are more tightly bound to the O$_{\mathrm{S}}$ atoms; see the peaks in the right-hand side panel of Fig.~\ref{fig1}B. 
	Their residence times at those groups is larger, so their mobility in this layer is limited. 
	Ca$^{2+}$ motion occurs mainly through hopping between binding sites, but such events are rare and contribute little to the total current.
	The region that contributes considerably to the total Ca$^{2+}$ current is the volume region in the middle of the pore.
	
	This statement is true for the anions as well but for a different reason.
	Cl$^{-}$ ions are effectively excluded from the surface region, so their surface conduction is also small. 
	In the volume region, Cl$^{-}$ velocity is larger due to their larger mobility (as in bulk), see the middle panel of Fig.~\ref{fig1}B. 
	As a consequence, the volume region, and thus, the pore as a whole, shows a bulk-like selectivity behavior for CaCl$_{2}$ as opposed to NaCl.
	
	For the ECCR2+TIP4P/2005 model, the ``bulk selectivity'' of CaCl$_{2}$ is $S_{+}^{\mathrm{B}}=0.713$.~\cite{salman_jml_2025} 
	In the scaled-charge O$_{\mathrm{S}}$ model, the pore selectivity is $S_{+}^{\mathrm{P}}=0.658$, while in the full-charge O$_{\mathrm{S}}$ model it is $0.446$  (these values correspond to $S_{+}^{*}=0.923$ and $0.626$ relative selectivities, respectively). 
	Thus, the negatively charged pore is even more Cl$^{-}$-selective than the bulk, and increasing the surface charge further enhances this trend.
	
	This effect arises from overcharging, a well-known phenomenon in multivalent electrolytes.~\cite{zheng_ep_2003,van_der_Heyden_prl_2006,li_aca_2019,li_nl_2015,lin_jacs_2020,siwy_prl_2002,siwy_cej_2006,he_jacs_2009,gillespie_bj_2008_nanopore,lorenz_pre_2007,Bourg_2011,hartkamp_pccp_2015,dopke_jpcc_2019,wang_jpcc_2021,fabian_jml_2022,rojano_pf_2024}
	The negatively charged surface attracts Ca$^{2+}$ ions so strongly that they overcompensate the wall charge, rendering the surface effectively positive.   
	Consequently, the monovalent Cl$^{-}$ ions behave with respect to this slightly positive region similarly as the Na$^{+}$ ions behave with respect to the original negative wall.
	If the effect is strong enough, an excess anion layer forms and a change in the sign of the electrical potential occurs; this phenomenon is coined as charge inversion.
	All these effects reflect strong ionic correlations beyond the mean-field Poisson-Boltzmann description.~\cite{boda_jcp_2002,matejczyk_jcp_2017}
	
	In summary, Ca$^{2+}$ ions do not have surface conduction due to their low \textit{mobility} in the surface region, while Cl$^{-}$ ions do not have surface conduction due to low \textit{availability}.
	This behavior is different from that of NaCl where Na$^{+}$ ions have a large surface conduction because they form a diffuse layer of more mobile ions than Ca$^{2+}$ ions do.
	This mechanism is clearly seen in the video clip uploaded to the \hyperref[si]{SM}.
	
\begin{figure*}[t!]
		\begin{center}
			\includegraphics*[width=0.65\textwidth]{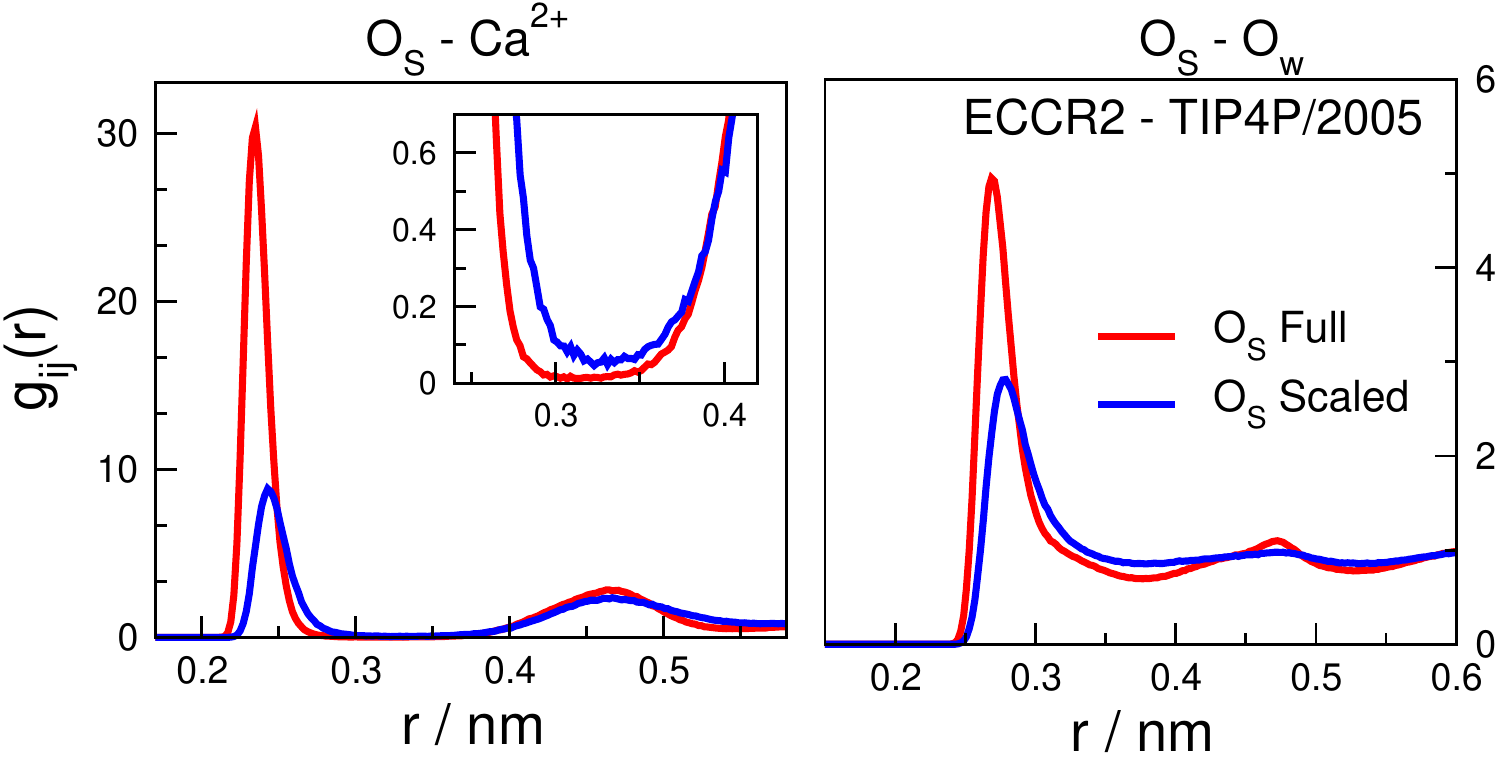}
		\end{center}
		\caption{Radial distribution functions (RDF) for pairs of O$_{\mathrm{S}}$ and Ca$^{2+}$ (left panel) as well as O$_{\mathrm{S}}$ and O$_{\mathrm{w}}$ (right panel), where O$_{\mathrm{S}}$ stands for the oxygen of the deprotonated silanol group and  O$_{\mathrm{w}}$ stands for the oxygen atom of the water molecule modeled by the TIP4P/2005 FF.
			The inset of the left panel zooms in on the depletion zone between the 1st and 2nd peaks.
			The figure shows results for the ECCR2 model with TIP4P/2005 water.
			Red and blue curves refer to full-charge and scaled-charge silanol oxygen (O$_{\mathrm{S}}$) models, respectively.}
		\label{fig2}
	\end{figure*}

	The tight binding of Ca$^{2+}$ ions to the O$_{\mathrm{S}}$ atoms is evident in Fig.~\ref{fig2} that shows RDFs of Ca$^{2+}$ (left panel) and the water oxygen (right panel) with respect to the O$_{\mathrm{S}}$ atom.
	The peaks are higher for the full-charge O$_{\mathrm{S}}$ model, indicating a practically non-ergodic behavior on the simulation time scale (i.e., slow dynamics), as seen from the very low values between the 1st and 2nd peaks in the inset of the left panel.
	
	This behavior was already observed in bulk simulations of full-charge models of Ca$^{2+}$ and Cl$^{-}$ (FULL and CHARMM).~\cite{salman_jml_2025} 
	In contrast, scaled-charge models exhibit physically more realistic transport properties, yielding diffusion coefficients and ionic conductivities in closer agreement with experimental data.~\cite{kohagen_jpcb_2015,martinek_jcp_2018,salman_jml_2025}
	
	Employing a scaled-charge representation for the deprotonated O$_{\mathrm{S}}$ atoms ensures consistency with the scaled-charge models used for the ions.
	The choice of the O$_{\mathrm{S}}$ model does not substantially alter the qualitative behavior of the pore.
	The reduced selectivity, relative to the bulk, is $S_{+}^{*} = 0.923$ for the scaled-charge pore and $0.626$ for the full-charge pore.
	Both values show a characteristic difference compared to the $S_{+}^{*}=2.39$ and $2.94$ values for NaCl.
	
	Our simulations, therefore, qualitatively agree with the experimental observations of He et al.~\cite{he_jacs_2009} for rectifying conical PET nanopores, where KCl showed cation selectivity, CoSepCl$_{3}$ (a 3:1 electrolyte) showed anion selectivity, and CaCl$_{2}$ displayed intermediate, nearly non-selective behavior.

	\subsection{The effect of pore radius}
	
	All simulations were performed at a bulk electrolyte concentration of $1$ M to reduce computational cost associated with the number of water molecules.
	To gain insight into how electrolytes of different bulk concentrations would behave in the pore of radius $R^{\mathrm{P}} \approx 1.3$ nm, it is more efficient to retain the $1$ M concentration and instead vary the pore radius so that the $\lambda_{\mathrm{D}}/R^{\mathrm{P}}$ ratio changes.
	We demonstrated in a series of papers that selectivity scales with this ratio, namely, concentrated electrolytes in a narrow pore behave similarly as dilute electrolytes behave in a wide pore.~\cite{sarkadi_jcp_2021,sarkadi_jml_2022,sarkadi_jml_2023} 
	
	While this approach has limitations, it minimizes changes in the total number of H$_{2}$O molecules in the simulation while allowing systematic exploration of confinement effects, the relation of the screening length and the pore radius, in this case.
	One major limitation is that a different concentration would result in charge inversion of different strength, and, thus, different extent of anion excess in the volume conduction region and different selectivity behavior. 
	We expect that this quantitative dependence does not influence the overall mechanism reported in this work.
	
	\begin{figure*}[t!]
		\begin{center}
			\includegraphics*[width=0.5\textwidth]{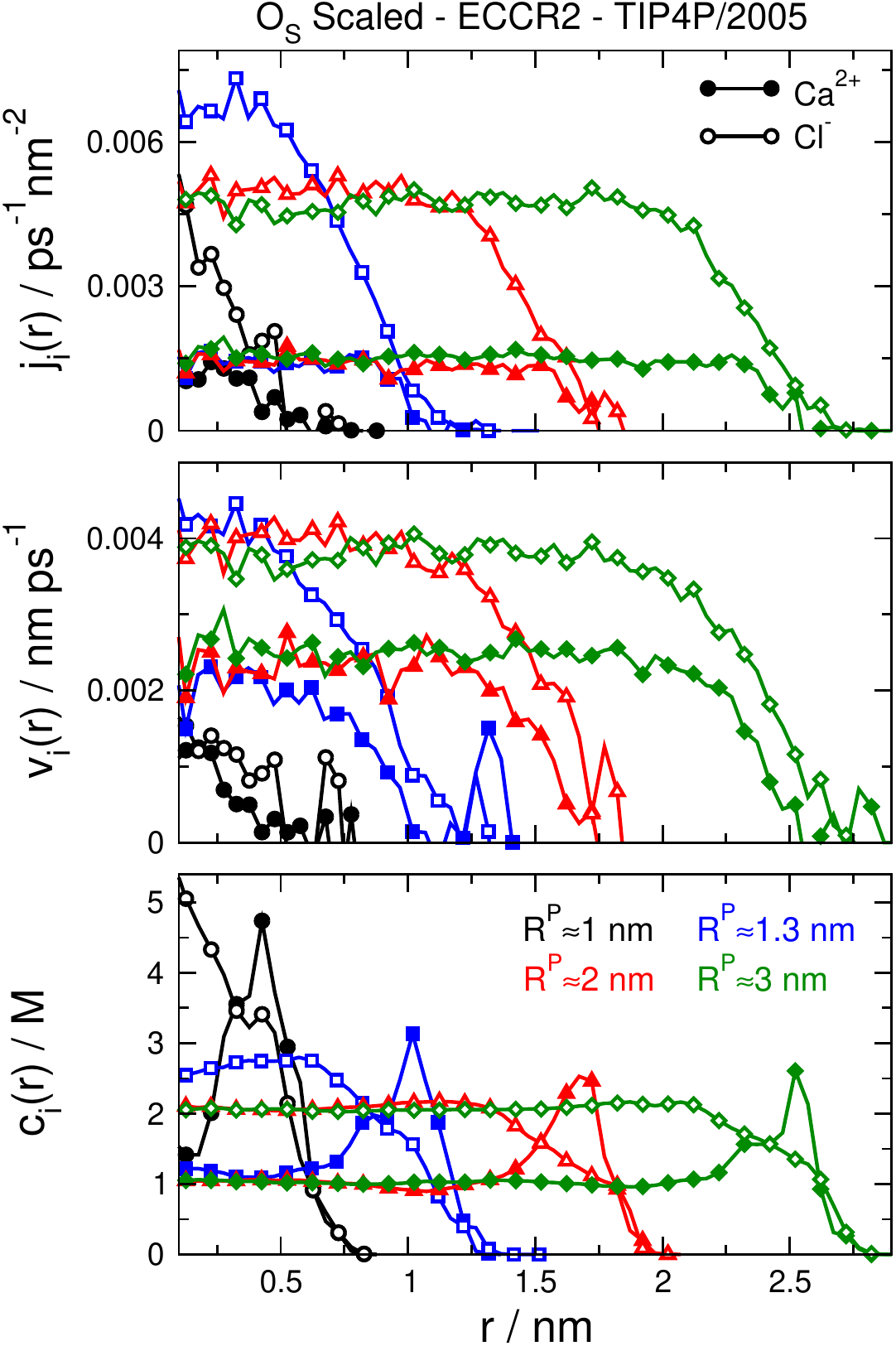}
		\end{center}
		\caption{From top to bottom: radial dependence of the axial ($z$) components of particle current density (in 1/ps$\;$nm$^{2}$), velocity (in nm/ps), and concentration (in mol/dm$^{3}$).
		Black, blue, red, and green curves refer to pore radii $R^{\mathrm{P}}\approx 1$, $1.3$, $2$, and $3$ nm, respectively.
		Full and open symbols refer to Ca$^{2+}$ and Cl$^{-}$ ions, respectively.
		The figure refers to simulations for the scaled-charge silanol oxygen (O$_{\mathrm{S}}$) model and the ECCR2 model of ions with TIP4P/2005 water.
		\label{fig3}}
	\end{figure*} 
	
	Figure~\ref{fig3} presents the $j_{i}(r)$, $v_{i}(r)$, and $c_{i}(r)$ profiles for different pore radii using the scaled-charge O$_{\mathrm{S}}$ model combined with the ECCR2 ion and TIP4P/2005 water models.
	As the pore radius increases, the bulk-like region at the center of the pore becomes wider.
	This region dominates over the surface region for the three larger radii for which the selectivities are approximately equal (see Table \ref{tab2}).
	Specifically, the reduced selectivities are $S_{+}^{*} \approx 0.923$, $0.978$, and $0.974$ for pore radii $R^{\mathrm{P}} \approx 1.3$, $2$, and $3$ nm, respectively.
	
	In the case of $R^{\mathrm{P}}\approx 1$ nm, however, the double layers overlap in the center of the pore (see the $c_{i}(r)$ profiles, black color).
	The electrolyte is less bulk-like in this pore, mobilities and current densities are smaller, and the $j_{i}(r)$ profiles for cations and anions are less separated.
	This results in a larger cation selectivity than in the case of larger radii.
	
	The selectivity mechanism described in this work, therefore, cannot be generalized to all nanopores or concentrations.
	For example, biological calcium channels are considerably narrower and possess a high density of structural charges (COO$^{-}$ groups from amino acid side chains), which not only exclude Cl$^{-}$ but also favor Ca$^{2+}$ over monovalent cations such as Na$^{+}$ and K$^{+}$.~\cite{almers_jp_1984a,boda_jcp_2006}
	
	\subsection{The effect of the ion model}
	
	While the above results were obtained using the ECCR2 ion model with TIP4P/2005 water, additional simulations with other ion and water models show that the overall ionic selectivity behavior described in the previous subsection is largely insensitive to the specific choice of the model.
	
\begin{figure*}[t!]
		\begin{center}
			\includegraphics*[width=0.9\textwidth]{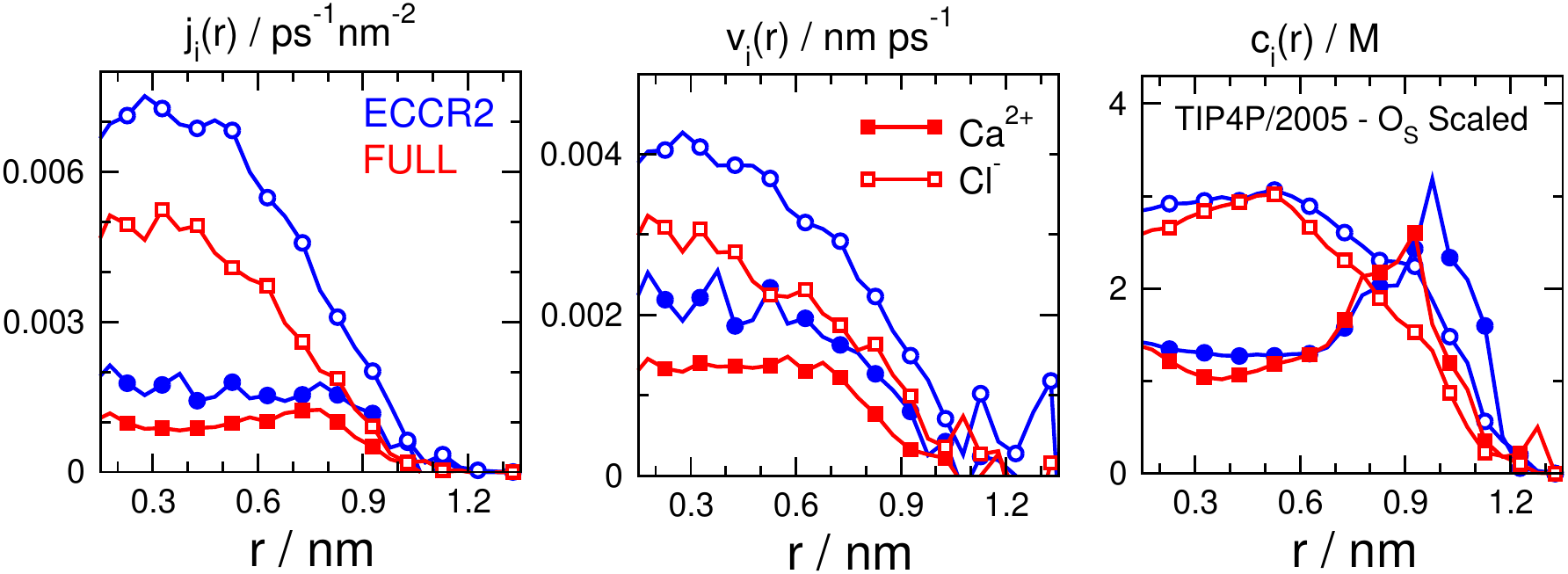}
		\end{center}
		\caption{From left to right: radial dependence of the axial ($z$) components of particle current density (in 1/ps$\;$nm$^{2}$), velocity (in nm/ps), and concentration (in mol/dm$^{3}$).
		Red and blue curves refer to the FULL and ECCR2 models of Ca$^{2+}$ and Cl$^{-}$, respectively.
		Full and open symbols refer to Ca$^{2+}$ and Cl$^{-}$ ions, respectively.
		The figure refers to simulations for the scaled-charge silanol oxygen (O$_{\mathrm{S}}$) model and the TIP4P/2005 model of water. }
		\label{fig4}
	\end{figure*}

	Figure \ref{fig4} shows the $j_{i}(r)$, $v_{i}(r)$, and $c_{i}(r)$ profiles for the full-charge ion model (FULL) in comparison with the scaled-charge ion model (ECCR2).
	Our bulk simulations \cite{salman_jml_2025} already indicated that Ca$^{2+}$ coordinates more strongly with water molecules than with Cl$^{-}$ in the FULL model compared to ECCR2.
	A similar situation arises here when O$_{\mathrm{S}}$ is considered to be the anion in this II+IW competition.
	The right panel of Fig.~\ref{fig4} shows that the peak of Ca$^{2+}$ near the wall is a bit farther from the wall for the FULL model (and also lower) than for the ECCR2 model.
	This is the effect of water molecules that bind stronger to the FULL Ca$^{2+}$ ions than to the ECCR2 Ca$^{2+}$ ions, thus favoring solvent-separated ion pairs.
	
	This influences the availability of Ca$^{2+}$ ions in the surface layer, where their mobility, on the other hand, is limited.
	In the bulk-like region, the conductivity and mobility of the ECCR2 ions are larger than those of the FULL ions due to their smaller charge and less tight water-shell.
	Pore selectivity is largely determined by the behavior in the bulk as shown by the very similar $S_{+}^{*}$ values ($0.923$ vs.\ $0.95$)
	
\begin{figure*}[t!]
		\begin{center}
			\includegraphics*[width=0.9\textwidth]{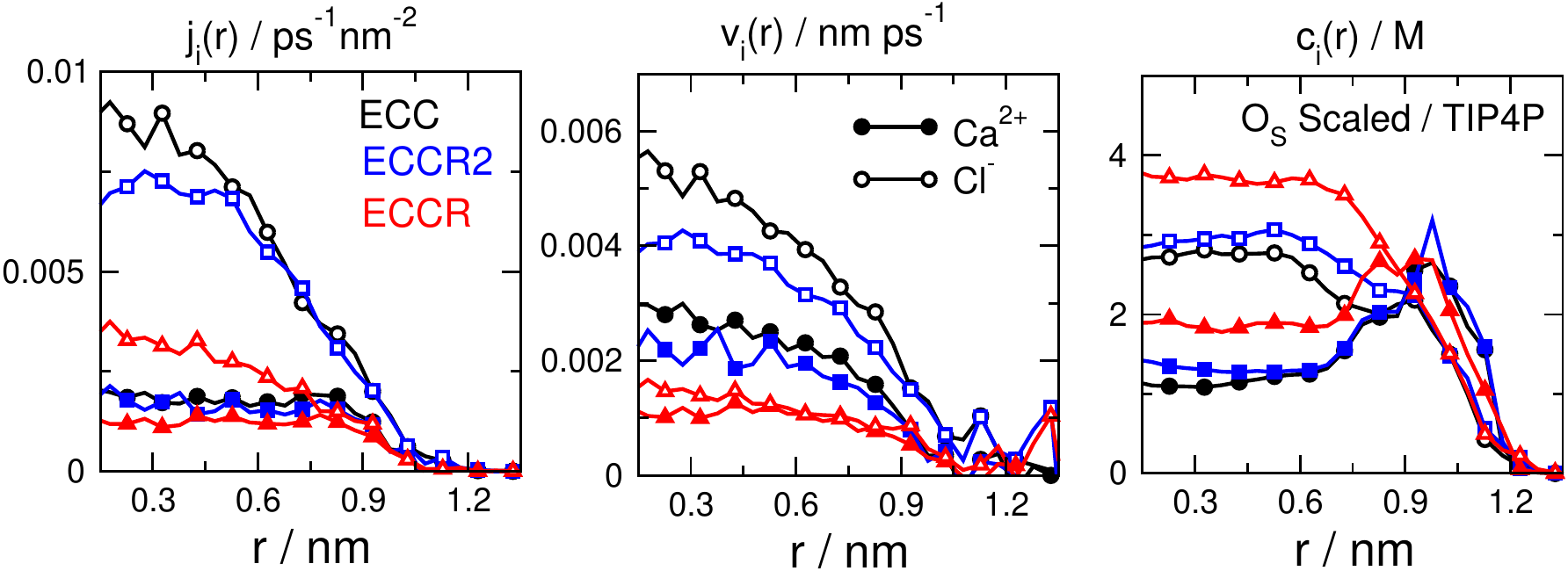}
		\end{center}
		\caption{From left to right: radial dependence of the axial ($z$) components of particle current density (in 1/ps$\;$nm$^{2}$), velocity (in nm/ps), and concentration (in mol/dm$^{3}$).
		Black, blue, and red curves refer to the ECC, ECCR2, and ECCR models of Ca$^{2+}$ and Cl$^{-}$, respectively.
		Full and open symbols refer to Ca$^{2+}$ and Cl$^{-}$ ions, respectively.
		The figure refers to simulations for the scaled-charge silanol oxygen (O$_{\mathrm{S}}$) model and the TIP4P/2005 model of water.}
		\label{fig5}
	\end{figure*}

	Figure \ref{fig5} shows the $j_{i}(r)$, $v_{i}(r)$, and $c_{i}(r)$ profiles for the scaled-charge ion models (ECC, ECCR, ECCR2) with scaled-charge O$_{\mathrm{S}}$ model and TIP4P/2005 water.
	While the overall behavior is the same, some discrepancy is observed in the case of ECCR, which is the scaled-charge ion  model with the smallest size.
	This model provides stronger binding between Ca$^{2+}$ and the O$_{\mathrm{S}}$ atom.
	The small Ca$^{2+}$ ion shows a structured behavior near the wall (a double peak is observed) and produces larger densities in the volume region.
	
	The anomaly lies rather in the behavior of the Cl$^{-}$ ions for the ECCR model. 
	The $v_{-}(r)$ profile indicates a reduced Cl$^{-}$ mobility compared to the other models.
	As a consequence, the $j_{-}(r)$ curve and Cl$^{-}$ current is also lower than for other models resulting in a slight cation selectivity.
	The explanation probably is that the ECCR Cl$^{-}$ ions associate strongly with the Ca$^{2+}$ ions as it was shown by our bulk simulations~\cite{salman_jml_2025}.
	
	It is a trend that smaller ions exhibit smaller mobilities (the ECC $\to$ ECCR2 $\to$ ECCR is the order of decreasing ion size) as shown by the velocity profiles (availability shows an opposite trend).
	This is in agreement with the results for the diffusion constant and conductivity in our bulk simulations~\cite{salman_jml_2025}.
	The explanation is that the tighter water shell around the smaller ions results in a larger hydrodynamic radius.
	
	The reduced selectivities are $S_{+}^{*}=0.941$, $0.923$, and $1.65$ for the ECC, ECCR2, and ECCR models, respectively, for the scaled-charge O$_{\mathrm{S}}$ pore.
	These numbers are $S_{+}^{*}=0.741$, $0.626$, and $0.82$ for the ECC, ECCR2, and ECCR models, respectively, for the full-charge O$_{\mathrm{S}}$ pore.
	
	The different behaviors of ECCR for the two pores ($S_{+}^{*}=1.65$ vs.\ $0.82$) are probably the consequence of the more complex competition between Ca$^{2+}$, Cl$^{-}$, water molecules, and O$_{\mathrm{S}}$ atoms.
	In bulk solutions, the O$_{\mathrm{S}}$ atoms were missing, so we were able to describe the behavior of the model on the basis of a competition of Cl$^{-}$ and H$_{2}$O at the Ca$^{2+}$ ions on the basis of the balance of II-IW (ion-ion vs.\ ion-water) interactions.
	The right balance prevented unphysical behavior such as too strong Ca$^{2+}$+Cl$^{-}$ association (II terms dominating) or too strong water shells around Ca$^{2+}$ (IW terms dominating). 
	The right balance also helped avoiding slow dynamics and practical non-ergodicity. 
	
	In the pore, association of Ca$^{2+}$ with O$_{\mathrm{S}}$ is influenced by the Cl$^{-}$ ions, but, in the meantime, Ca$^{2+}$+Cl$^{-}$ pair formation is influenced by the O$_{\mathrm{S}}$ atoms.
	The role of water molecules in this picture might seem secondary.
	This is, however, not the case; the water model has a serious influence on the system's behavior.
	
	\subsection{The effect of the water model}
	
\begin{figure*}[t!]
		\begin{center}
			\includegraphics*[width=0.9\textwidth]{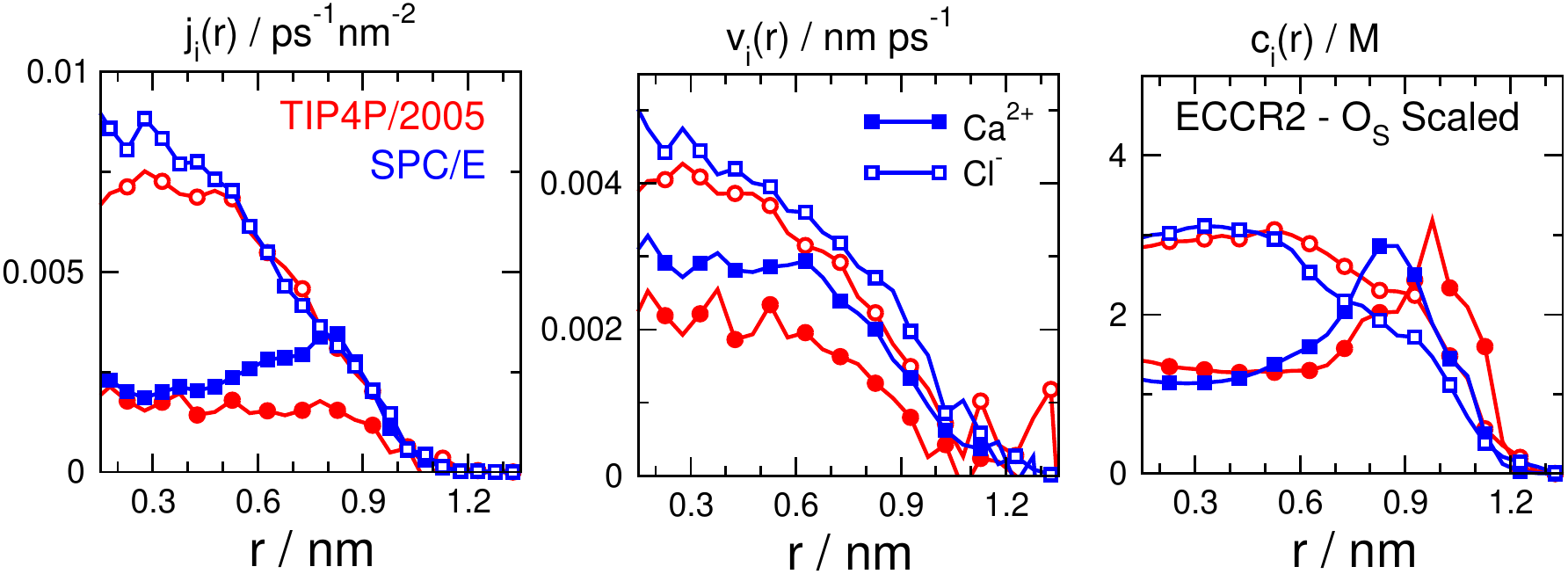}
		\end{center}
		\caption{From left to right: radial dependence of the axial ($z$) components of particle current density (in 1/ps$\;$nm$^{2}$), velocity (in nm/ps), and concentration (in mol/dm$^{3}$).
		Red and blue curves refer to the TIP4P/2005 and SPC/E models of water, respectively. 
		Full and open symbols refer to Ca$^{2+}$ and Cl$^{-}$ ions, respectively.
		The figure refers to simulations for the scaled-charge silanol oxygen (O$_{\mathrm{S}}$) model and the ECCR2 model of ions.}
		\label{fig6}
	\end{figure*}

	Figure \ref{fig6} shows the $j_{i}(r)$, $v_{i}(r)$, and $c_{i}(r)$ profiles for  TIP4P/2005 and SPC/E water models with the ECCR2 ion model and the scaled-charge O$_{\mathrm{S}}$ model being fixed.
	
	It was already apparent from our results for bulk that the SPC/E water model is stickier than the TIP4P/2005 model: it interacts with the ions more strongly and forms a tighter water shell around the ions.
	The consequence of this fact in the nanopore is twofold.
	We observe a Ca$^{2+}$ peak farther from the wall for the SPC/E model due to the SPC/E molecules associating with Ca$^{2+}$ more strongly and hindering the Ca$^{2+}$+O$_{\mathrm{S}}$ association. 
	This makes the Ca$^{2+}$ ions more available in a region, where they are more mobile due to their increased distance from the binding sites.
	The increased mobility of the ions is also shown by the velocity profiles; it is also a consequence of the reduced (more screened by water) interactions between the charged species (Ca$^{2+}$, Cl$^{-}$, and O$_{\mathrm{S}}$).
	
\begin{figure*}[t!]
		\begin{center}
			\includegraphics*[width=0.65\textwidth]{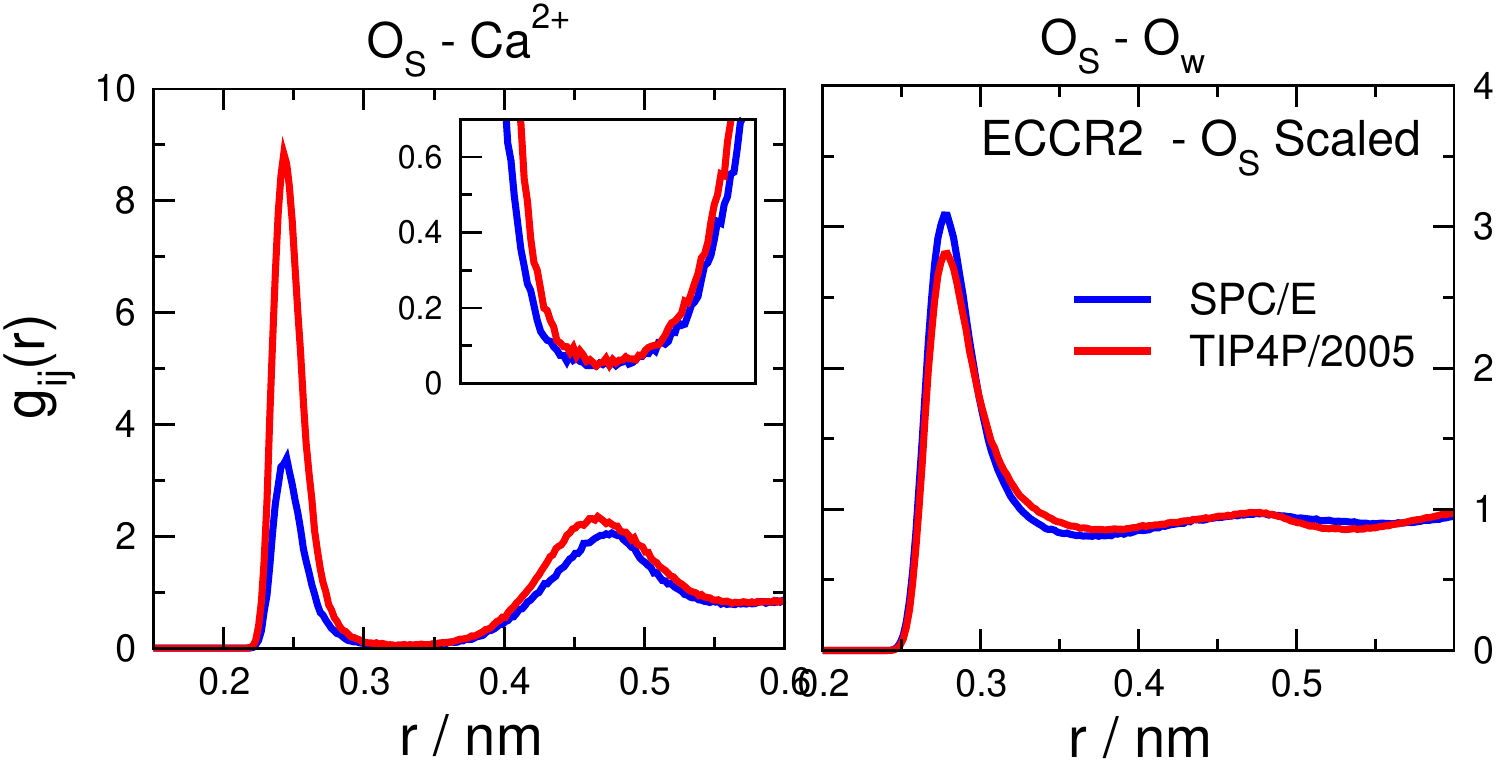}
		\end{center}
		\caption{Radial distribution functions for pairs of O$_{\mathrm{S}}$ and Ca$^{2+}$ (left panel) as well as O$_{\mathrm{S}}$ and O$_{\mathrm{w}}$ (right panel), where O$_{\mathrm{S}}$ stands for the oxygen of the deprotonated silanol group and  O$_{\mathrm{w}}$ stands for the oxygen atom of the water molecule.
		Red and blue curves refer to the TIP4P/2005 and SPC/E models of water, respectively. 
		The inset of the left panel zooms in on the depletion zone between the 1st and 2nd peaks.
		The figure shows results for the ECCR2 model with TIP4P/2005 water.
		The figure refers to simulations for the scaled-charge silanol oxygen (O$_{\mathrm{S}}$) model and the ECCR2 model of ions.}
		\label{fig7}
	\end{figure*}

	The fact that the competition between Ca$^{2+}$ and water for the space near the O$_{\mathrm{S}}$ atoms is different for the two water models is well shown by the RDF profiles in Fig.~\ref{fig7}.
	The 1st peak in the O$_{\mathrm{S}}$-Ca$^{2+}$ RDF is higher for the TIP4P/2005 model, while the reverse behavior is observed for the O$_{\mathrm{S}}$-O$_{\mathrm{w}}$ RDF. 
	The O$_{\mathrm{S}}$-Ca$^{2+}$ curve for SPC/E (left panel, blue color) is not something that we expect for the RDF between two oppositely charged particles. 
	The 1st peak is too small, indicating a weak binding of Ca$^{2+}$ ions to the O$_{\mathrm{S}}$ atoms hindered by the SPC/E water molecules.
	This has consequences for the current and selectivity data.
	
	If we look at the pore selectivity for SPC/E with the ECCR2 ion model, it is $S_{+}^{\mathrm{P}}=1.05$ for the scaled-charge O$_{\mathrm{S}}$ pore, while it is $S_{+}^{\mathrm{P}}=0.84$ for the full-charge O$_{\mathrm{S}}$ pore.
	These numbers are systematically larger then the corresponding values for TIP4P/2005: $S_{+}^{\mathrm{P}}=0.664$ and $0.447$.
	
	For SPC/E, the reduced selectivity values are $S_{+}^{*}=2.54$ ($0.923$ for TIP4P/2005) for the scaled-charge O$_{\mathrm{S}}$, and  $S_{+}^{*}=1.293$ ($0.626$ for TIP4P/2005) for the full-charge O$_{\mathrm{S}}$.
	This is the result of SPC/E water molecules hindering the Ca$^{2+}$+O$_{\mathrm{S}}$ interactions. 
	This favors Ca$^{2+}$ selectivity.
	However, the resulting $S_{+}^{*}$ numbers are still close to 1, so the pore's selectivity behavior is still bulk-like.
	
	Summarized, Ca$^{2+}$ selectivity is more sensitive to the choice of water model (cation selectivities are systematically larger for SPC/E) and the charge of the O$_{\mathrm{S}}$ group (cation selectivities are systematically larger for the scaled-charge O$_{\mathrm{S}}$ model) than to the choice of the ion model.

	\subsection{Electroosmotic flow}
	
	\begin{figure*}[t!]
		\begin{center}
			\includegraphics*[width=0.45\textwidth]{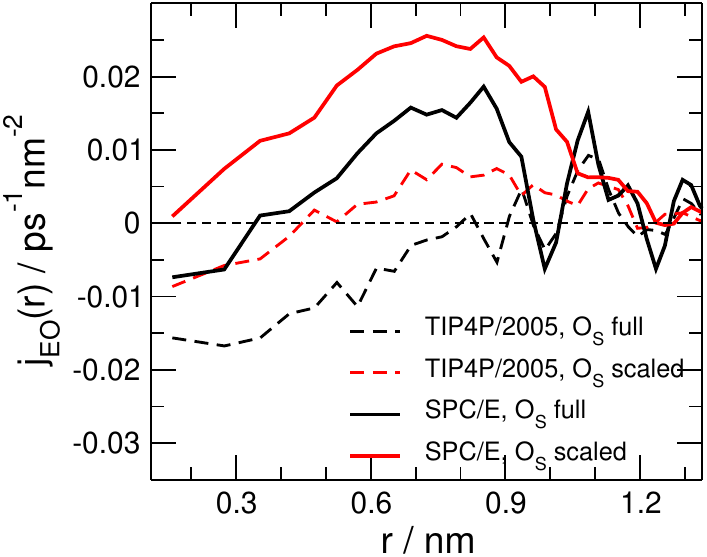}
		\end{center}
		\caption{Radial electroosmotic current density profiles $j_{\mathrm{EO}}(r)$ for the ECCR2 ion model combined with different pore and water models.}
		\label{fig8}
	\end{figure*}

	Although the primary focus of this study is ionic transport, our simulations also provide access to the dynamics of water molecules.
	Electroosmotic flow (EOF) arises from the momentum transfer between moving ions and the surrounding solvent: ions that dominantly carry electrical current drag water molecules along with them and produce a net EOF.
	Water transport is therefore a secondary effect of coupled ion–solvent motion; the applied electric field does not act directly on water molecules (note that no pressure gradient is imposed).
	
	Because the pore is effectively non-selective, no single ionic species acts as a dominant charge carrier throughout the entire cross-section.
	As a result, the direction and magnitude of the EOF are governed by a delicate interplay of two factors:
	(i) the relative contributions of the different ionic species in distinct regions of the pore (surface versus volume), and
	(ii) the strength of their coupling to water molecules.
	Consequently, the sign of the net water velocity depends sensitively on the chosen ion, pore, and water models.
	In most of our simulations, the average velocity of water is positive, corresponding to flow in the direction of Ca$^{2+}$ migration.
	An exception is observed for the full-charge O$_{\mathrm{S}}$ pore model combined with TIP4P/2005 water, where the net flow reverses sign, which is generally observed in experiments~\cite{van_der_Heyden_prl_2006} due to charge inversion in CaCl$_2$.
	The associated uncertainties are relatively large, reflecting the weak electric field strength employed in the simulations.
	
	To analyze electroosmotic behavior locally, Fig.~\ref{fig8} presents radial electroosmotic current density profiles $j_{\mathrm{EO}}(r)$ for the ECCR2 ion model combined with all pore and water model variants.
	In the surface region near the pore wall, Ca$^{2+}$ ions dominate the ionic population.
	However, their motion occurs primarily through rare hopping events between binding sites, resulting in limited sampling and substantial statistical uncertainty.
	Although the corresponding $j_{\mathrm{EO}}(r)$ values in this region are generally small and positive, their detailed behavior cannot be resolved reliably.
	
	In contrast, more definitive conclusions can be drawn for the central, more bulk-like region of the pore.
	For all four model combinations, $j_{\mathrm{EO}}(r)$ increases as the wall is approached.
	This trend indicates that, in the pore center, water molecules preferentially move with Cl$^{-}$ ions, whereas closer to the charged wall, they are more strongly coupled to Ca$^{2+}$ ions.
	Three of the four profiles even change sign at intermediate radial positions, although at different values of $r$.
	As a result, water flows in opposite directions in different regions of the pore, leading to a net EOF that is small in magnitude and uncertain in sign.
	
	Systematic shifts between the profiles reflect differences in model choices.
	The solid curves corresponding to SPC/E water are consistently more positive than the dashed curves obtained with TIP4P/2005.
	This behavior is attributed to the stronger ion–water coupling with SPC/E, which has been shown to be effectively more ``sticky''~\cite{salman_jml_2025}, leading to a higher probability of water molecules traveling with Ca$^{2+}$ ions.
	Similarly, the red curves associated with the scaled-charge O$_{\mathrm{S}}$ pore model are more positive than the black curves corresponding to the full-charge model.
	This shift reflects the larger Ca$^{2+}$ flux in the scaled-charge case (Fig.~\ref{fig1}B), resulting from weaker ion–surface attraction.
	
	These findings are in harmony with results reported in previous MD studies. 
	Our results highlight the strong sensitivity of EOF to model parameters, consistent with numerous reports in which the direction and magnitude of EOF depend sensitively on simulation conditions. For example, Hartkamp et al.~\cite{hartkamp_pccp_2015} reported negative EOF even for monovalent electrolytes. Předota et al.~\cite{predota_l_2016} found negative EOF for Na$^{+}$ and Sr$^{2+}$, but positive EOF for Rb$^{+}$ at negatively charged surfaces. Rojano et al.~\cite{rojano_pf_2024} showed that the addition of even a small amount of MgCl$_2$ to NaCl can reverse the EOF direction. Together, these studies demonstrate that EOF emerges from a delicate balance of multiple effects, including local ionic excess, ion mobility, ion–surface interactions, ion–water coupling, and viscosity.
	
	Further evidence of this sensitivity is provided by Rezaei et al.~\cite{Rezaei_2018}, who showed that the EOF velocity exhibits a maximum as a function of surface charge density: beyond this maximum, increasing surface charge hinders cation mobility and reduces its contribution to positive EOF.
	
	Notably, all of these MD studies employed full-charge ion models and reported their findings without a systematic assessment of how FF choices influence electrokinetic predictions. 
	The present subsection is not intended to address this issue comprehensively; rather, it aims to provide limited insight into the sensitivity of electrokinetic behavior to modeling choices. 
	We suggest that further systematic studies of FF transferability in electrokinetic simulations are needed, and we hope that the results presented here help motivate such investigations.
	
	\section{Conclusions}
	
	In this work, we used atomistic MD simulations to investigate ion transport through negatively charged silica nanopores in NaCl and CaCl$_2$ solutions. 
	Our results show that the transport mechanisms in these two electrolytes are qualitatively different. 
	While NaCl exhibits conventional cation-selective behavior governed by mobile counterions in the EDL, CaCl$_2$ displays a loss of cation selectivity associated with strong Ca$^{2+}$ adsorption, partial immobilization near the pore surface, and dominant anion transport in the pore interior following charge inversion.
	
	In this paper and its preceding study \cite{salman_jml_2025}, we discuss the sensitivity to modeling choices in different contexts.
	(1) In bulk, we found that ion and water models influence the balance of II and IW interactions, resulting in observable differences in experimentally measurable transport properties.~\cite{salman_jml_2025}
	(2) When these electrolyte models are placed in a silica nanopore, a robust qualitative behavior of ions emerges, modulated by the strong attractive effect of surface oxygens. This behavior is largely insensitive to modeling choices, although quantitative differences remain.
	(3) As a secondary effect, these quantitative differences lead to small changes in the EOF that, although minor, can result in a change in the sign of the EOF, thus underscoring the sensitivity to modeling choices.
	
	Despite the growing number of molecular simulations addressing electrokinetic transport, systematic investigations of FF dependence remain scarce, particularly for confined systems such as nanopores. 
	Our results highlight that quantities of electrokinetic phenomena, including ionic currents and EOF, are sensitive to modeling choices, emphasizing the need for careful validation and comparison of FFs in nanoscale confinement.
	
	Finally, although the detailed transport behavior depends on a delicate balance between ion–ion, ion–surface, and ion–water interactions, the qualitative mechanisms identified here are consistent across the models considered.
	This robustness implies that the overall principle of ion selectivity is governed by ``important'' degrees of freedom (ionic charges, ionic sizes, surface group modeling), while ``less important'' degrees of freedom (water modeling and chemical details of the pore wall beyond the charged groups) can be replaced by simplified representations.
	These simplified representations include implicit models of water and hard-wall confinements as in our studies employing reduced models for electrolytes~\cite{vincze_jcp_2010,valisko_fpe_2023}, EDLs~\cite{boda_jcp_2002,henderson_pccp_2009}, ion channels~\cite{boda_jcp_2006,boda_bj_2007,gillespie_bj_2008_ca,boda_jgp_2009,malasics_bba_2010_trivalent}, and nanopores~\cite{gillespie_bj_2008_nanopore,valisko_jcp_2019,boda_entropy_2020,sarkadi_jcp_2021,sarkadi_jml_2022}.  
	These studies demonstrated that reduced models may capture the essential physics of a system to be able to reproduce device-level (input-output) behavior.
	
	\section*{Supplementary Material}
		\label{si}
		The SM consists of a video clip showing the different conduction mechanisms for NaCl and CaCl$_2$ and a document that shows results for electric field strength dependence, simulation time dependence, and a figure showing reduced selectivity data collected in Table \ref{tab2}.
		
	\section*{Acknowledgements}
	
	This work has been implemented by the National Multidisciplinary Laboratory for Climate Change (RRF-2.3.1-21-2022-00014) project within the framework of Hungary's National Recovery and Resilience Plan supported by the Recovery and Resilience Facility of the European Union.
	We gratefully acknowledge  the financial support of the National Research, Development and Innovation Office -- NKFIH K124353 and TKP2021-
	NKTA-21.
	We acknowledge KIFÜ (Governmental Agency for IT Development, Hungary, https://ror.org/01s0v4q65) for awarding us access to the Komondor HPC facility based in Hungary.

	\appendix
	\section{Determining the radial profiles}
	\label{appendix}
	
	To obtain the radial profiles, we average Eqs.~\ref{eq:vialpha}-\ref{eq:jialpha} over the $\phi$ and $z$ coordinates.
	Let us rewrite Eq.~\ref{eq:jvc1} as 
	\begin{equation}
		j_{i} dV = v_{i}dN_{i}
	\end{equation} 
	whose discretized form is
	\begin{equation}
		j_{i}^{\alpha} V^{\alpha} = v_{i}^{\alpha} \frac{N_{i}^{\alpha}}{N_{t}}.
	\end{equation} 
	We sum over $N_{v}$ volume elements as  
	\begin{equation}
		\sum_{\alpha=1}^{N_{v}} j_{i}^{\alpha} V^{\alpha} = \frac{1}{N_{t}} \sum_{\alpha=1}^{N_{v}} v_{i}^{\alpha} N_{i}^{\alpha}
	\end{equation}
	The left-hand side is
	\begin{equation}
		V^{\mathrm{tot}} \left( \frac{1}{V^{\mathrm{tot}}} \sum_{\alpha=1}^{N_{v}} j_{i}^{\alpha} V^{\alpha} \right) = V^{\mathrm{tot}} \bar{j}_{i} ,
		\label{eq:left}
	\end{equation}
	where $\bar{j}_{i}$ is defined by the expression in the parentheses as the average current density averaged over the total volume $V^{\mathrm{tot}}$.
	Using Eq.~\ref{eq:vialpha} for $v_{i}^{\alpha}$, the right hand side can be expressed as
	\begin{align}
		\frac{1}{N_{t}} \sum_{\alpha=1}^{N_{v}} \left( \dfrac{1}{N_{i}^{\alpha}}\sum_{k=1}^{N_{i}^{\alpha}}\dfrac{\Delta z_{i,k}}{\Delta t} \right) N_{i}^{\alpha} = \nonumber \\
		\frac{1}{N_{t}} \sum_{\alpha=1}^{N_{v}} \sum_{k=1}^{N_{i}^{\alpha}}\dfrac{\Delta z_{i,k}}{\Delta t} = \nonumber \\
		\frac{N_{i}^{\mathrm{tot}}}{N_{t}} \left( \frac{1}{N_{i}^{\mathrm{tot}}} \sum_{k=1}^{N_{i}^{\mathrm{tot}}} \dfrac{\Delta z_{i,k}}{\Delta t} \right) = \bar{N}_{i} \bar{v}_{i} ,
		\label{eq:right}
	\end{align}
	where $N_{i}^{\mathrm{tot}}$ is the number of ions found in the total volume $V^{\mathrm{tot}}$ in the sampled configurations, so $\bar{N}_{i}=N_{i}^{\mathrm{tot}}/V^{\mathrm{tot}}$ is the average number of ions in the total volume.
	From the equality of Eqs.~\ref{eq:left} and \ref{eq:right} we obtain that
	\begin{equation}
		\bar{j}_{i}=\frac{\bar{N}_{i}}{V^{\mathrm{tot}}} \bar{v}_{i} = \bar{c}_{i}\bar{v}_{i},
	\end{equation}
	where $\bar{c}_{i}$ is the average concentration in the total volume. 
	This derivation shows that space averaging and the multiplication $cv$ is interchangable in the case of averaging over particle movements in Eq.~\ref{eq:vialpha}.
	The radial profiles are computed by fixing the radial distance $r^{\alpha}=\sqrt{(r_{\mathrm{o}}^{\alpha})^{2} - (r_{\mathrm{i}}^{\alpha})^{2}}$ and summing over $N_{\phi}$ and $N_{z}$ elements both  in the azimuthal and axial dimensions ($N_{v}=N_{\phi}N_{z}$).
	Note that if the volume $V^{\alpha}$ of the elementary cells is constant (as is the case in our study), the average in Eq.~\ref{eq:left} is reduced to a simple mean average.

	\bibliography{nanopore,book,own}
	\bibliographystyle{unsrt} 
	
\end{document}